\documentclass[prb,twocolumn,notitlepage,superscriptaddress]{revtex4-1}
\usepackage[T1]{fontenc}
\usepackage[utf8]{inputenc}

\usepackage{graphicx}
\usepackage{amsmath,amssymb,amsfonts,mathtools}
\usepackage{xcolor}
\usepackage{tikz}
\usetikzlibrary{shapes.geometric,arrows,decorations.markings}
\usepackage{enumitem}
\usepackage{siunitx}
\newcommand{\Bip}{B_\mathrm{ip}}
\newcommand{\Boop}{B_\mathrm{oop}}
\newcommand{\Phib}{\Phi_\mathrm{b}}
\DeclareSIUnit{\pg}{pg}
\DeclareSIUnit{\GHz}{GHz}
\usepackage[normalem]{ulem}

\usepackage{microtype}

\usepackage[colorlinks=true,citecolor=magenta,linkcolor=blue,urlcolor=blue]{hyperref}

\begin{document}

\title{Resolving Abrikosov vortex entry in superconducting nano-string resonators via displacement-noise spectroscopy in cavity-optomechanics}

\author{Thomas Luschmann}

\affiliation{Walther-Meißner-Institut, Bayerische Akademie der Wissenschaften, 85748 Garching, Germany}
\affiliation{Physics Department, TUM School of Natural Sciences, Technical University of Munich, 85748 Garching, Germany}

\author{Tahereh Sadat Parvini}
\email{tahereh.parvini@wmi.badw.de}
\affiliation{Walther-Meißner-Institut, Bayerische Akademie der Wissenschaften, 85748 Garching, Germany}

\author{Lukas Niekamp}
\affiliation{Walther-Meißner-Institut, Bayerische Akademie der Wissenschaften, 85748 Garching, Germany}
\affiliation{Physics Department, TUM School of Natural Sciences, Technical University of Munich, 85748 Garching, Germany}

\author{Achim Marx}
\affiliation{Walther-Meißner-Institut, Bayerische Akademie der Wissenschaften, 85748 Garching, Germany}

\author{Rudolf Gross}
\affiliation{Walther-Meißner-Institut, Bayerische Akademie der Wissenschaften, 85748 Garching, Germany}
\affiliation{Physics Department, TUM School of Natural Sciences, Technical University of Munich, 85748 Garching, Germany}
\affiliation{Munich Center for Quantum Science and Technology (MCQST), 80799 Munich, Germany}

\author{Hans Huebl}
\email{huebl@wmi.badw.de}
\affiliation{Walther-Meißner-Institut, Bayerische Akademie der Wissenschaften, 85748 Garching, Germany}
\affiliation{Physics Department, TUM School of Natural Sciences, Technical University of Munich, 85748 Garching, Germany}
\affiliation{Munich Center for Quantum Science and Technology (MCQST), 80799 Munich, Germany}

\date{\today}
\begin{abstract}

Abrikosov vortices in type-II superconductors critically influence current flow and coherence, thereby imposing fundamental limits on superconducting quantum technologies. Quantum circuits employ superconducting elements at micro- and mesoscopic scales, where individual vortices can significantly impact device performance, necessitating investigation of vortex entry, motion, and pinning in these constrained geometries. Cavity-optomechanical platforms combining flux-tunable microwave resonators with superconducting nanomechanical elements offer a promising route to the single-photon strong-coupling regime and enable highly sensitive probing of the mechanical degree of freedom under elevated magnetic fields. Here, we exploit this platform to investigate vortex entry processes at the single-event level. We observe discrete jumps of the mechanical resonance frequency attributable to individual vortex entry,  corresponding to attonewton-scale forces and allowing quantitative extraction of single-vortex pinning energies. These signatures are superimposed on a smooth power-law background characteristic of the collective Campbell-regime of vortex elasticity. Our results establish optomechanics-inspired sensing as a powerful method for exploring fundamental superconducting properties and identifying decoherence pathways in quantum circuits. Beyond advancing vortex physics, this work opens new opportunities for integrating mechanical sensing into superconducting device architectures, bridging condensed matter physics and quantum information science.

\end{abstract}

\maketitle

\section{Introduction}
Nanomechanical resonators have emerged as powerful sensors for the detection of minuscule forces, accelerations, or mass \cite{Moser:2013go, fogliano2021ultrasensitive, yang2006zeptogram, jensen2008atomic, cai2025room, tao2014single, tavernarakis2018optomechanics, tabatabaei2024large, eom2011, Biswas:2014ku, hirose2025coupled,  chaste2012}. In addition, they also allow the investigation of the intrinsic properties of the underlying materials such as Young's modulus, stress, loss mechanisms, and coupling between elastic and other degrees of freedom in the solid-state \cite{Hocke:2014jx, weber2016, klass2022determining, li2023tuning, bereyhi2022hierarchical, vsivskins2025nonlinear, mathew2015nanoscale, zhang2013nanomechanical, kanellopulos2025comparative, chen2009performance, staruch2019magnetoelastic, karabalin2009piezoelectric, bhaskar2016flexoelectric, meesala2016enhanced, Pernpeintner:2016bm,schwienbacher2019}. Often, these properties are encoded in the displacement and resonance frequency of the oscillator, which are then measured using optical, electrical, or opto-mechanical schemes \cite{Metzger.2004, Schwab:2005fy, AspelmeyerRMP}. Cavity optomechanics, also known as nano-electromechanics, combines superconducting (quantum) microwave circuits with mechanical oscillators and has the potential to study mechanical resonators made of superconducting materials.
\begin{figure}[!t]
    \centering
    \includegraphics[width=0.49\textwidth]{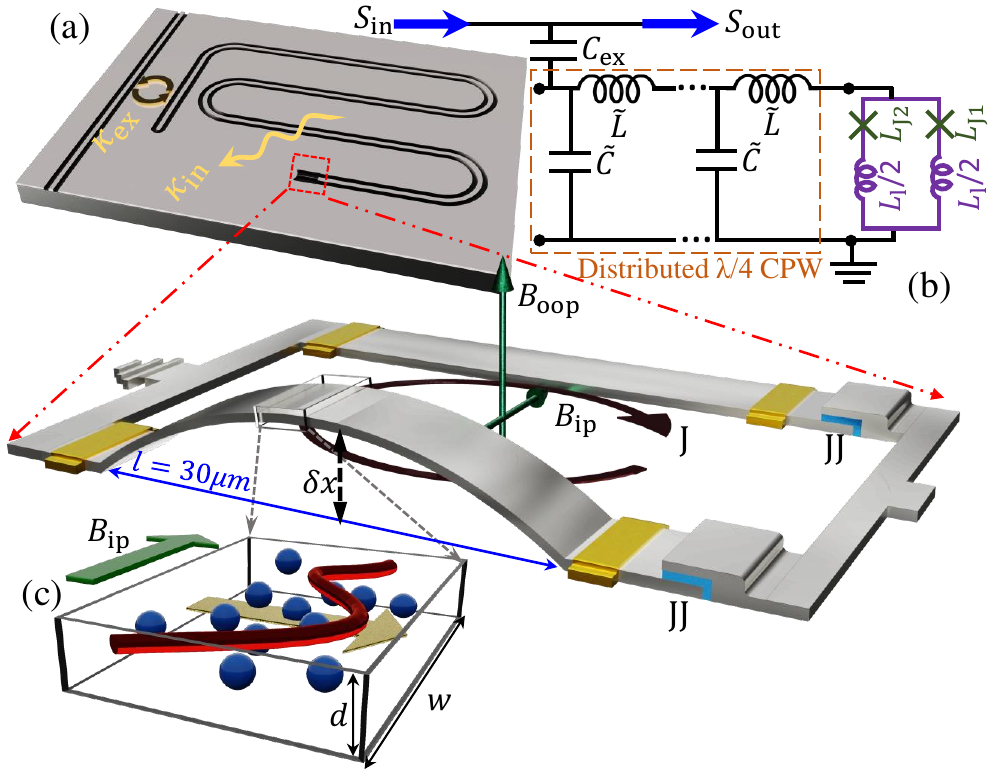}
    \caption{\textbf{Nanomechanical detection of magnetic flux lines.} (a) Superconducting \(\lambda/4\) coplanar waveguide resonator capacitively coupled to a microwave feedline and terminated to ground by a dc SQUID. $\kappa_\mathrm{ext}$ and $\kappa_\mathrm{in}$ denote the external and internal coupling rates of the resonator, respectively. The magnified view shows the SQUID loop incorporating a suspended Al nanostring \((l,w,d)=(30,\,0.2,\,0.11)\,\si{\micro\meter})\) and two Josephson junctions (JJs), with labels indicating the nanostring displacement \(\delta x\), circulating SQUID current \(J\), and applied magnetic control fields \(\Bip\) (in-plane) and \(\Boop\) (out-of-plane). (b) Lumped-element circuit representation of the distributed $\lambda/4$ CPW resonator with distributed inductance \(\tilde{L}\) and capacitance \(\tilde{C}\) per unit length, external coupling capacitor \(C_\mathrm{ex}\) and the feedline ports ($S_\mathrm{in}$ and $S_\mathrm{out}$). The total resonator inductance and capacitance are denoted as $L$ and $C$, respectively. The SQUID provides flux-tunable termination through geometric loop inductance (\(L_\mathrm{l}\), distributed as \(L_1/2\) on each branch) and Josephson inductances (\(L_{J1}(\Phi)\) and \(L_{J2}(\Phi)\)). (c) Conceptual depiction of magnetic flux penetration into thin-film Al nanostring, where vortices become trapped at pinning centers (blue spheres).}
    \label{Fig1}
\end{figure}

Type-II superconductors exposed to magnetic fields above the lower critical field $H_\mathrm{c1}$ enter the mixed state, where magnetic flux lines penetrate the material. The pinning and motion of these vortices govern the critical current and generate dissipation in the flux-flow state~\cite{tinkham2004}. Mesoscopic structures are particularly suited for studying individual vortex behavior, as their  dimensions enable the resolution of discrete vortex nucleation, spatial positioning, and pinning dynamics. However, conventional electrical transport measurements only probe the ensemble-averaged behavior, while local imaging techniques can be perturbative and often require specialized cryogenic instrumentations~\cite{Geim-nature:2000, Stan:2004, Stoddart.1993, foltyn2023,Foltyn:2024}. This motivates the development of integrated, minimally invasive single-vortex detection schemes. Historically, vortex dynamics have been studied through mechanical properties using vibrating reed techniques~\cite{Brandt1987, esquinazi1991, brandt1986, gupta1991, kober1991vibrating}. Recently, the emergence of cavity optomechanics with superconducting circuits~\cite{rodrigues2019, schmidt2020, Bothner.2022, luschmann2022mechanical} has enabled unprecedented sensitivity for probing flux dynamics in nanoscale superconductors. Nanomechanical resonators with cavity readout combine high force sensitivity with continuous, non-invasive monitoring~\cite{AspelmeyerRMP, Rabl:2011gn, buks2007, blencowe2007, schmidt2020}, making them well-suited for detecting individual vortex events. This raises the question whether mechanical resonators with sub-micron-sized diameters allow for the direct detection of individual vortex nucleation and depinning events, field-driven vortex-array rearrangements, and pinning potentials.

In this work, we employ an inductively coupled cavity–optomechanical (or nano-electromechanical) platform to study frequency variations of a superconducting string resonator as a function of an applied magnetic field. Our experiments suggest that the observed frequency changes originate from the entry of individual magnetic flux lines into the string resonator with a sub-micron diameter.  

\section{Device Concept}
\begin{figure}[t]
    \centering
    \begin{minipage}[b]{0.54\columnwidth}
        \raisebox{-0.5cm}{
        \includegraphics[width=4.86cm, height=2.1cm]{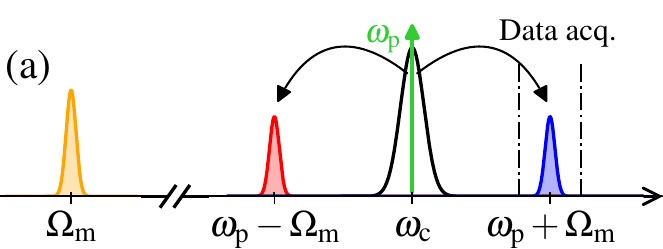}}
    \end{minipage}
    \hfill
    \begin{minipage}[b]{0.44\columnwidth}
        \centering
        \raisebox{-1.3cm}{
        \includegraphics[width=3.6cm, height=2.9cm]{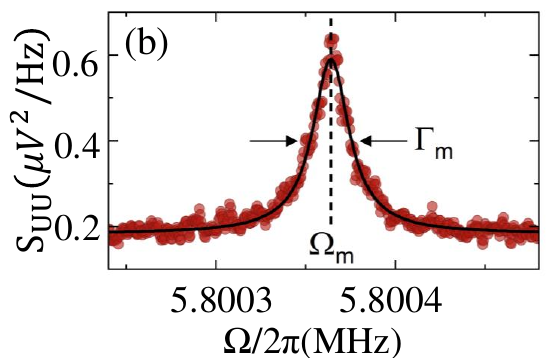}}
    \end{minipage}
    
    \vspace{-0.7cm}
    
    \begin{minipage}[b]{\columnwidth}
        \centering
        \includegraphics[width=\textwidth]{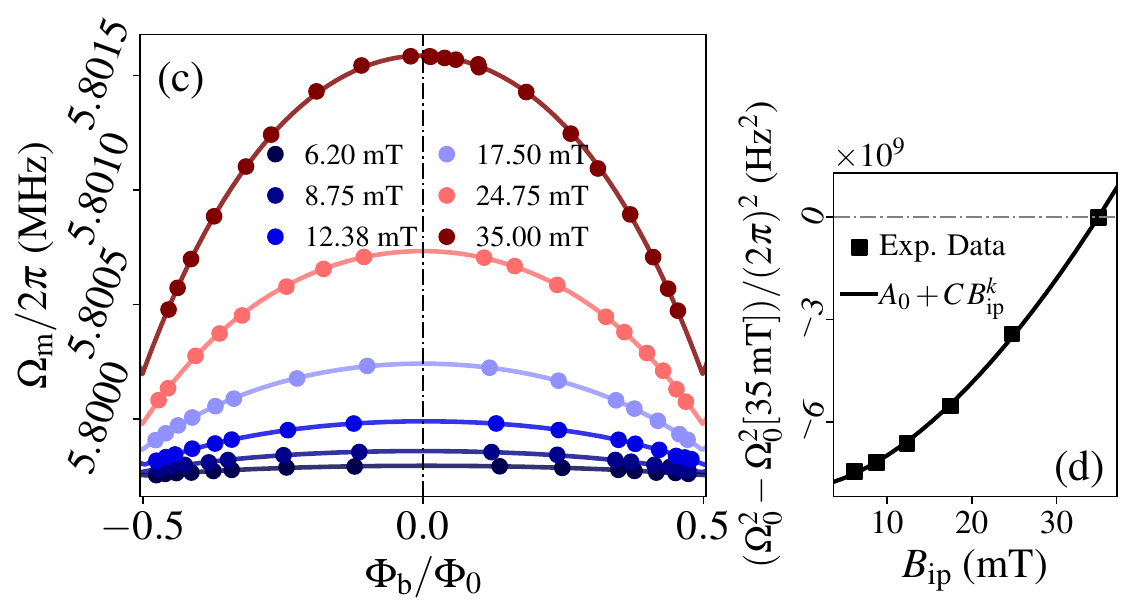}
    \end{minipage}
        
    \caption{\textbf{Measurement scheme for determining the resonance frequency of the string-resonator.}  (a) Sketch of all relevant modes of the cavity optomechanical system. The mechanical mode at $\Omega_\mathrm{m}$ interacts with the probe tone $\omega_\mathrm{p}$ resulting in two mechanical sidebands at $\omega_\mathrm{p} \pm \Omega_\mathrm{m}$. For the determination of the thermal displacement noise, used to infer $\Omega_\mathrm{m}$, we record the anti-Stokes field for $\omega_\mathrm{p}=\omega_\mathrm{c}$. (b) Representative voltage power spectral density of the anti-Stokes field downconverted to $\omega_\mathrm{p}$ showing mechanical resonance, from which we determine the mechanical resonance frequency $\Omega_\mathrm{m}$ and linewidth $\Gamma_\mathrm{m}/2\pi=\SI{20}{Hz}$. (c) Evolution of the mechanical frequency $\Omega_\mathrm{m}(\Phib/\Phi_0)$ as a function of $\Boop$ measured for different $\Bip$. (d) Extracted intrinsic mechanical frequency $\Omega_0(\Bip)$ versus in-plane magnetic field, obtained by fitting the flux-dependent data to the theoretical model (Eq.\,\ref{fre}). The solid line is a power-law fit $A_0+C\,B_{\mathrm{ip}}^{\,k}$ with exponent \(k=1.81\pm0.05\).}
    \label{Fig2}
\end{figure}

We realize a hybrid nano-electromechanical platform in which a superconducting aluminum (Al) nanostring resonator is integrated into the loop of a flux-sensitive dc SQUID. The SQUID terminates the end of a superconducting $\lambda/4$ coplanar waveguide (CPW) resonator to ground, thereby realizing a flux-tunable optomechanical interaction~\cite{rodrigues2019, schmidt2020, luschmann2022mechanical}. The device is fabricated using standard nanofabrication techniques, combining double-layer Al shadow evaporation with reactive-ion etching on a high-resistivity silicon substrate. The suspended nanostring has dimensions $(l,w,d)=(30,0.2,0.11)\,\si{\micro\meter}$ and an effective mass of $m_\mathrm{eff}=\SI{0.6}{pg}$ (for details, see Ref.~\cite{schmidt2020}). Figure~\ref{Fig1}(a) depicts the device layout, with an inset highlighting the dc SQUID containing the two Josephson junctions (JJs) and the suspended nanostring, together with the applied in-plane ($\Bip$) and out-of-plane ($\Boop$) magnetic control fields and the nanostring displacement $\delta x$. The equivalent circuit is shown in Fig.~\ref{Fig1}(b). The CPW resonator has a resonance frequency of $\omega_\mathrm{c}/2\pi\approx\SI{7.4}{GHz}$ at zero flux bias. The dc SQUID’s flux-tunable inductance enables in-situ control of the CPW resonance from \SI{7.4}{GHz} to \SI{6.6}{GHz} via $\Boop$~(see Ref.~\cite{luschmann2022mechanical}). In addition, the motion of the nanostring in the magnetic field modulates the flux threading the SQUID loop, thereby changing the SQUID’s effective inductance. Hereby, the displacement $\delta x$ affects the resonance frequency of the microwave resonator and implements an optomechanical interaction with a single-photon optomechanical coupling rate  $g_0=\eta\,(\partial_\Phi \omega_\mathrm{c}) \,l\, B\,x_\mathrm{zpf}$. Here, $\eta$, $l$, and $B$ represent the mode shape factor, the length of the nanostring, and the applied magnetic flux density, respectively~\cite{buks2007, blencowe2007, nation2016, rodrigues2019, schmidt2020}. In our experiment, we control flux-responsivity of the microwave resonator  $\partial_\Phi \omega_\mathrm{c} =\partial \omega_\mathrm{c} / \partial\Phi$ by application of an out-of-plane bias field $\Boop$. The superposition with an additional in-plane flux density $\Bip$ results in a total static $B$ and enables significant enhancement of the optomechanical coupling rate $g_0$ due to the enhanced field resilience of thin-film superconductors against in-plane magnetic fields~\footnote{We assume that, in our device geometry, any field enhancement from Mei{\ss}ner-Ochsenfeld screening in the thin-film superconducting regions is negligible.}. The zero-point motion is $x_\mathrm{zpf}=\sqrt{\hbar/(2 m_\mathrm{eff}\Omega_\mathrm{m})}\approx \SI{50}{fm}$, where $\Omega_\mathrm{m}/2\pi=\SI{5.8}{MHz}$ is the mechanical resonance frequency.

\section{Experimental and Theoretical Considerations}

Measurements are performed in a dilution refrigerator at $T \approx \SI{85}{\milli\kelvin}$ after cooling down the sample at zero applied magnetic field (ZFC: zero-field-cooled). Details of the microwave setup and magnetic field control are given in Refs.~\cite{luschmann2022mechanical, schmidt2020}. For mechanical spectroscopy, we use thermal noise sideband spectroscopy with an ultra-low-power microwave probe tone at frequency $\omega_\mathrm{p}$ ($P_\mathrm{probe}<\SI{2}{\femto\watt}$) tuned into resonance with the flux-tunable cavity, $\omega_\mathrm{p}=\omega_\mathrm{c}(\Phi)$. The optomechanical interaction generates motional sidebands at $\omega_\mathrm{p}\pm\Omega_\mathrm{m}$ (see Fig.~\ref{Fig2}(a)). Operating in the linearized regime, the optomechanical coupling is enhanced to $g=g_0\sqrt{n_\mathrm{c}}$, where $n_\mathrm{c}$ is the intracavity photon number. Under our experimental conditions, the cooperativity $C = 4g^2/(\kappa\Gamma_\mathrm{m})$ remains below unity, placing the system in the weak-coupling regime where optomechanical backaction is negligible. Here, $\kappa = \kappa_\mathrm{in} + \kappa_\mathrm{ext}$ ($\kappa/2\pi \approx \SI{2.5}{MHz}$) denotes the total cavity decay rate, and $\Gamma_\mathrm{m}$ is the mechanical damping rate. Nevertheless, the large thermal occupation of the mechanical mode even at millikelvin temperatures imprints resolvable motional sidebands on the cavity response~\cite{AspelmeyerRMP}. The probe power is kept sufficiently low to ensure that the flux-tunable resonator operates in its linear response regime~\cite{Bothner.2022, schmidt2020, Dhiman.2025}. The anti-Stokes sideband at $\omega_\mathrm{p}+\Omega_\mathrm{m}$ directly encodes the mechanical response (Fig.~\ref{Fig2}(b)), from which we extract $\Omega_\mathrm{m}$ by linking the voltage power spectral density to the mechanical susceptibility
\begin{equation}
  \chi_\mathrm{m}^{-1}(\Omega) = m_\mathrm{eff}\,\left[\Omega_\mathrm{m}^2 - \Omega^2 - i\,\Gamma_\mathrm{m}\Omega\right].
\label{eq:susc_mech}
\end{equation}
Complementary vector network analyzer (VNA) measurements are used to extract the flux-tunable mechanical frequency in regimes where the anti-Stokes sideband signal is too weak to resolve. Sweeping the flux bias $\Phib$ at a fixed in-plane field $\Bip$ by varying $\Boop$ produces a periodic, dome-shaped modulation of $\Omega_\mathrm{m}(\Phib)$ (see Fig.~\ref{Fig2}(c)). This modulation dominantly arises from flux-dependent circulating currents in the SQUID loop (denoted by $J$ in Fig.~\ref{Fig1}(a)). In the presence of the in-plane field, a Lorentz force acts on these currents and, in turn, on the nanostring motion~\cite{shevchuk2017,rodrigues2019,luschmann2022mechanical}. To quantify the observed $\Omega_\mathrm{m}(\Phib;\Bip)$, we model the nanostring as a damped harmonic oscillator driven by a displacement-dependent Lorentz force~\cite{rodrigues2019,winkel2020implementation}. The Lorentz force $F_\mathrm{L}=\eta l\Bip J$ arises from the interaction of $\Bip$ with the circulating current $J$ and introduces a flux-dependent modification of the mechanical frequency. The equation of motion reads
\begin{equation}
    \ddot{\delta x} + \Gamma_\mathrm{m}\dot{\delta x} + \Omega_0^2 \delta x = \frac{F_\mathrm{L}}{m_\mathrm{eff}},
    \label{derivat}
\end{equation}
where $\Omega_0$ is the intrinsic angular frequency of the unperturbed mechanical mode. The resulting effective mechanical frequency is \cite{rodrigues2019}
\begin{equation} 
\Omega_\mathrm{m}^2(\Phib;\Bip) = \Omega_0^2(\Bip) + \frac{\eta^2 \Bip^2 l^2}{m_\mathrm{eff}\,[2L_\mathrm{J}(\Phib)+L_\mathrm{l}]},
\label{fre}
\end{equation}
where $L_\mathrm{l}$ is the geometric loop inductance and $L_\mathrm{J}(\Phib)=\Phi_0/[2\pi I_{\mathrm{c}}|\cos(\pi\Phib/\Phi_0)|]$ is the Josephson inductance of a single junction with critical current $I_\mathrm{c}$ and $\Phi_0=h/(2e)$ is the magnetic flux quantum. Fitting the measured $\Omega_\mathrm{m}(\Phib)$ data with Eq.~\eqref{fre} yields $\Omega_0(\Bip)$, as shown in Fig.~\ref{Fig2}(d). Note that $\Omega_0(\Bip)$ corresponds to the baseline frequency obtained after subtraction of the Lorentz force induced spring term (second term in Eq.~\eqref{fre}), which can be directly observed as $\Omega_\mathrm{m}(\Phib \to (n+\tfrac{1}{2})\Phi_0)$ where the Josephson inductances diverge \footnote{We assume symmetric Josephson junctions.}. In contrast, at $\Phib=0$ the string is maximally stiffened and therefore $\Omega_\mathrm{m}(\Phib=0)$ exceeds $\Omega_0(\Bip)$. The fitting procedure and initial parameter values are provided in Appendix~\ref{app:Fitting}.

\begin{figure*}[t!]
  \centering
  \includegraphics[width=0.8\textwidth]{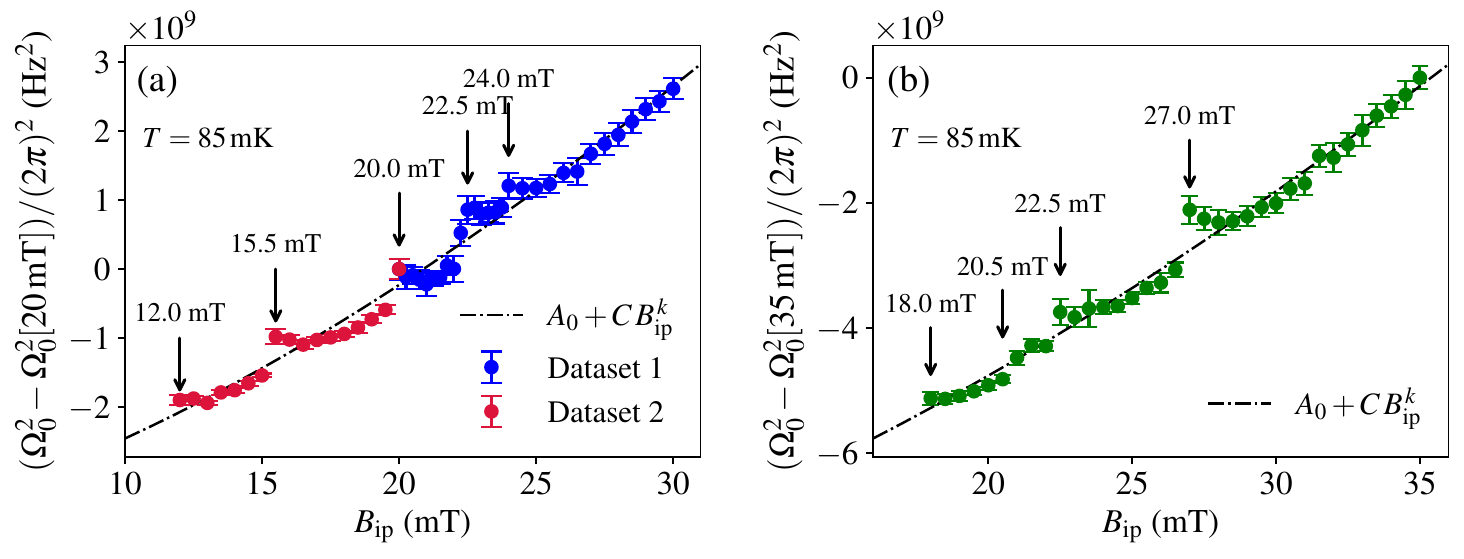}
    \caption{\textbf{Stochastic vortex entry signatures across independent cooldowns.} Squared mechanical frequency shift versus in-plane magnetic field at 85 mK for (a) two successive datasets within the same cooldown (blue and red circles), referenced to $\Omega_0(\SI{20}{mT})$ and (b) an independent dataset taken after warming to room temperature and re-cooling, referenced to $\Omega_0(\SI{35}{mT})$. The dash-dot line shows the power-law fit $A_0 + CB_{\mathrm{ip}}^k$ with $k=1.81\pm0.05$. Arrows mark abrupt frequency steps attributed to vortex entry events.}
  \label{Fig3}
\end{figure*}

Extending our previous work~\cite{luschmann2022mechanical}, we increased the in-plane field resolution to \SI{0.25}{\milli\tesla}, enabling high-resolution characterization of the mechanical response. Following the protocol presented in Fig.~\ref{Fig2}(a–d), we extract $\Omega_0(\Bip)$ with high precision, as shown in Fig.~\ref{Fig3}(a,b). Two measurement runs were performed during the same cooldown (Fig.~\ref{Fig3}(a)), and a third after a complete thermal cycle (Fig.~\ref{Fig3}(b)). These measurements reveal that $\Omega_0^2(\Bip)$ follows an offset power law,
\begin{equation}
    \Omega_0^2(\Bip) = \Omega_{0,\mathrm{bare}}^2 + \vartheta\,\Bip^{k},
    \label{offset}
\end{equation}
where $\Omega_{0,\mathrm{bare}}$ is the zero-field resonance frequency and $\vartheta, k$ quantify the field-induced renormalization. Although the smooth, near-quadratic background could arise from magnetomotive stiffening, in-plane screening, or higher-order corrections to the Lorentz-force term in Eq.~\eqref{fre}, the discrete, step-like discontinuities superimposed on this trend - whose positions vary between cooldowns - are incompatible with such smooth, reversible effects. Consistent with prior experiments~\cite{bolle1999observation}, the stochastic, history-dependent jumps in $\Omega_0(\Bip)$ are the hallmark of vortex entry and pinning in thin-film type-II superconductors, placing these observations beyond the vortex-free electromagnetic-spring model of Eq.~\eqref{derivat}. We explore this interpretation in detail in the following sections.

\section{Vortex Entry Signatures in the Resonance of an Aluminum Nanostring} 

To justify a vortex-based interpretation, we first quantify the superconducting parameters of the Al film under our conditions. The ability of a superconductor to host vortices is governed by the Ginzburg–Landau parameter $\kappa_\mathrm{GL}=\lambda/\xi$, the ratio of the magnetic penetration depth $\lambda$ to superconducting coherence length $\xi$. Bulk crystalline Al, with $\lambda_\mathrm{L}^{\infty} = \SI{50}{nm}$ and $\xi^{\infty}_\mathrm{GL} = \SI{1600}{nm}$, has $\kappa_\mathrm{GL}^\infty = 0.031 < 1/\sqrt{2}$ and is therefore a type-I superconductor~\cite{GrossMarx, cyrot1973ginzburg, Alloul2011, pippard1953, lopez2023magnetic}. However, thin films can deviate strongly from bulk behavior~\cite{meservey1971, Moshchalkov.1995, tinkham2004}. Our polycrystalline Al film has a grain size of $\sim\SI{10}{nm}$. Using this as an estimate for the electron mean free path  places the nanostring in the dirty limit ($l_\mathrm{f} \approx \SI{10}{nm} \ll \xi_\mathrm{GL}^{\infty}$). In this regime, disorder renormalizes the characteristic length scales as~\cite{brandt1971, de2018superconductivity}:
\begin{align}
    \lambda &= \alpha\,\lambda_\mathrm{L}^\infty \sqrt{\frac{\xi_\mathrm{GL}^\infty}{l_\mathrm{f}}}, 
    \qquad  
    \xi \simeq 0.85\,\sqrt{\xi_\mathrm{GL}^\infty\,l_\mathrm{f}},
    \label{xilambda}
\end{align}
where $\alpha=1.33$ accounts for surface scattering in thin films~\cite{tinkham2004, lopez2023magnetic}. Substituting our parameters yields $\xi \approx \SI{108}{nm}$ and $\lambda \approx \SI{0.84}{\micro\meter}$. Notably, $\xi$ is comparable with the thickness and width of our nanostring, while $\lambda>w,d$. In ultra thin polycrystalline films, boundary scattering and granularity can renormalize these characteristic lengths, so the quoted values should be regarded as estimates. Nevertheless, with $\kappa_\mathrm{GL}=\lambda/\xi \approx 7.8 > 1/\sqrt{2}$, the nanostring operates as a type-II superconductor capable of hosting vortices~\cite{tinkham2004, brandt1971, buckel2008superconductivity}. 

In the thin-film, narrow-strip limit, edge screening sheet currents concentrate near the boundaries, lowering surface barriers and enabling edge-mediated entry of vortices aligned with the in-plane field at weak nucleation sites set by polycrystalline disorder and edge roughness~\cite{Bronson:2006, Song:2009, Foltyn.2023, stan2004, Geim-nature:2000, kimmel2019edge}. The first-entry field follows from the balance of vortex line (self) energy and the work done by these edge currents. Under field-cooled (FC) conditions, the screening established during cooling can enhance the Bean–Livingston barrier, delaying penetration to $B_\mathrm{c1}^{\mathrm{FC}}=1.65\,\Phi_0/d^2 \approx\SI{282}{mT}$~\cite{roitman2024suppression}. Under ZFC conditions, the first-entry field is ~\cite{stan2004, likharev1971formation} 
\begin{equation}
  B_\mathrm{c1}^{\mathrm{ZFC}} =\frac{\Phi_0}{2\pi\xi d }\approx\SI{27.7}{\milli\tesla},
\end{equation} 
which we regard as an upper bound applicable to an ideal, homogeneous film with smooth edges. Bronson et al.~\cite{Bronson:2006, Milosevic.2010} showed that disorder and edge roughness suppress the surface barrier and reduce the vortex-entry field. In the nanostring, a principal mechanism for the observed near-quadratic dependence of $\Omega_0^2(\Bip)$ is progressive vortex penetration. The trend sets in at $\Bip\approx\SI{5}{mT}$ (Fig.~\ref{Fig2}(d)). We therefore take $B_\mathrm{c1}\approx\SI{5}{mT}$ as the empirical vortex-entry field for this nanostring. For $\Bip > B_\mathrm{c1}$, the vortex density in the nanostring can be estimated following Bronson et al.~\cite{Bronson:2006}:
\begin{equation}
    n_v \equiv \frac{N}{ld} = \frac{\Bip - \sqrt{\Bip B_\mathrm{c1}}}{\Phi_{0}},
\end{equation}
where $N$ is the number of vortices penetrating the nanostring. Near threshold ($\Bip\gtrsim B_\mathrm{c1}$), this reduces to $n_v \approx (\Bip- B_\mathrm{c1})/(2\Phi_0)$, implying discrete field steps of $\Delta B = 2\Phi_0/(ld) \approx \SI{1.25}{mT}$ per vortex. At higher fields, where $\Bip\gg B_\mathrm{c1}$, the density approaches the uniform limit $n_v \to \Bip/\Phi_0$, and the spacing reduces to $\Delta B \approx \Phi_0/(ld) \approx \SI{0.63}{mT}$. Notably, local thickness decrease at rough edges increases $\Delta B$. However, Fig.~\ref{Fig3} reveals irregular jump spacings of \SIrange{1.5}{5}{\milli\tesla}. Following Ref.~\cite{bolle1999observation}, we attribute these features to discrete vortex entry, with each jump signaling penetration of one or a few flux quanta at localized defects.

\section{Vortex-induced elastic response of the nanostring} 

Vibrating type-II superconducting resonators exhibit magnetic-field-dependent resonance shifts through several well-established mechanisms~\cite{brandt1986, kober1991vibrating}. In the mixed state ($B_{c1}<\Bip<B_{c2}$), Abrikosov vortices (flux lines) penetrate the nanostring one by one, with geometry-dependent entry pathways that modify the resonance frequency. One such mechanism is magnetoelastic coupling between Abrikosov vortices and the crystal lattice, established within the Ginzburg–Landau theoretical framework~\cite{Labusch.1968, Miranovic:1992, Kogan:1995}. The physical origin of this coupling is the so-called volume effect, that is, the fact that the specific volumes of the normal and superconducting phases are different. Therefore, the normal cores of flux lines result in local deformations in the surrounding superconducting material, allowing vortices to interact via the resulting elastic field~\cite{Miranovic:1992, Kogan:1995}. Since the relative volume change of Al is of the order of $10^{-8}$~\cite{ott1972}, the expected frequency shift in our tensile-strained nanostrings is negligibly small, $\Delta\Omega_0/\Omega_0 \sim 10^{-8}$, and cannot explain the observed discrete steps of magnitude $\Delta\Omega_0/\Omega_0 \sim 10^{-5}$. While the sign of the effect, an increase in $\Omega_0$, is consistent with normal-state contraction, the magnitude rules out this mechanism as the dominant contribution.

\begin{figure}[t!]
\includegraphics[width=0.47\textwidth]{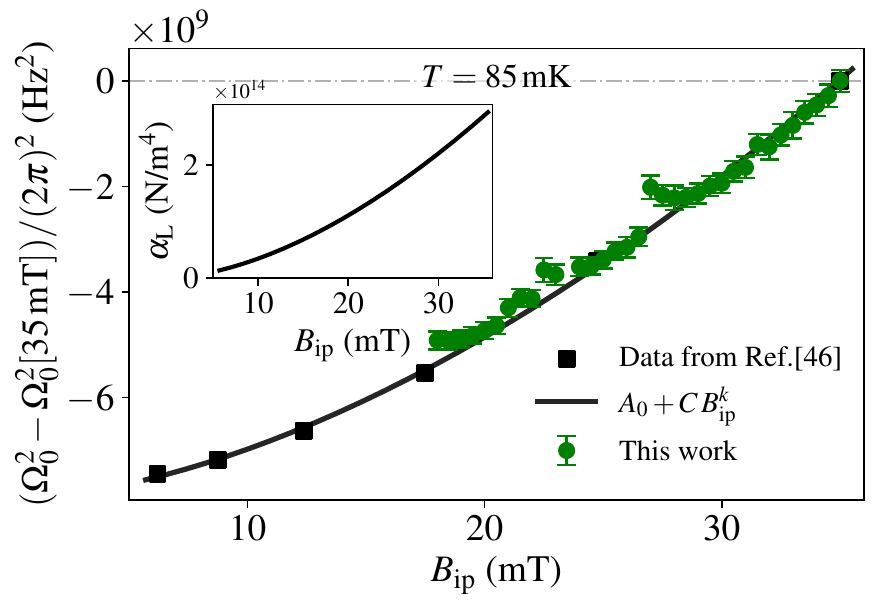}
	\caption{\textbf{Collective vortex stiffening and extracted Labusch parameter.} Squared resonance frequency shift (relative to 35 mT) versus in-plane field for high-resolution data (blue circles with error bars) and previous measurements~\cite{luschmann2022mechanical} (black squares). The black line represents the fitting function with $ k=1.81\pm 0.05$. Inset shows extracted Labusch parameter $\alpha_L(B_{\mathrm{ip}})$ as a function of $B_{\mathrm{ip}}$.}
	\label{Fig4}
\end{figure}

The dominant mechanism governing these frequency shifts is vortex pinning elasticity. In the small-amplitude Campbell regime, vortices trapped at defects or sample edges oscillate within their pinning wells, acting as distributed springs that stiffen the resonator. Microscopically, a small transverse displacement $r$ of a pinned vortex from equilibrium generates a linear restoring force per unit length \(f_{\mathrm{pin}}\approx -k_p\,r\), with the single-vortex pinning stiffness \(k_p\equiv\left.\partial_r^2 U(r)\right|_{r=0}\) for a local pinning potential \(U(r)\). The Campbell/Labusch stiffness is the ensemble-averaged curvature of the effective pinning potential per unit vortex length, $\alpha_{\mathrm{L}}\equiv\langle \partial_r^2 U_{\mathrm{eff}}(r)\rangle$~\cite{Labusch.1968, brandt1986}, and depends statistically on the vortex configuration. In the uniform-pinning approximation $\alpha_{\mathrm{L}}\equiv n_{v}\langle k_p\rangle\simeq (\Bip/\Phi_0)\langle k_p\rangle$, in which pinned vortices add a uniform spring density to the nanostring resonator~\cite{esquinazi1991, roitman2024suppression, foltyn2023probing, ge2023vortex, mironov2017anomalous}. For the (nearly) uniform fundamental mode of the nanostring, this shifts the eigenfrequency as
\begin{equation}
  \Omega_0^2(\Bip) = \Omega_{0,\mathrm{bare}}^2 + \frac{\alpha_\mathrm{L}(\Bip,T)}{\rho_\mathrm{eff}}\,,
  \label{eq:labusch_model}
\end{equation}
where $\rho_{\mathrm{eff}} = m_{\mathrm{eff}}/V_{\mathrm{eff}}$ is the modal effective mass density (for our device, $\rho_{\mathrm{eff}}=910\,\mathrm{kg\,m^{-3}}$). The bulk Labusch parameter typically exhibits a power-law field dependence $\alpha_\mathrm{L}(\Bip)\propto \Bip^k$, where the exponent $k$ reflects pinning strength and the elastic response of the vortex ensemble~\cite{kober1991, brandt1995flux, blatter1994vortices}. The nanostring’s baseline frequency exhibits robust power-law scaling with $k=1.81\pm0.05$, reproduced consistently across both previous low-resolution data~\cite{luschmann2022mechanical}(black filled circles, Fig.~\ref{Fig4}) and the present high-resolution measurements, shown with error bars in Figs.~\ref{Fig3} and \ref{Fig4}. The extracted exponent is consistent with the Campbell regime in strongly pinned, disordered type-II superconductors~\cite{esquinazi1991, kober1991, gupta1991}, where collective elastic stiffness arises from disorder-induced pinning. Interpreted in this way, the Labusch stiffness $\alpha_{\mathrm{L}}$, extracted from $\Delta\Omega_0$ (inset of Fig.~\ref{Fig4}), spans $(0.1\text{--}2)\times10^{14}\,\mathrm{N\,m^{-4}}$, consistent with macro-scale vibrating-reed measurements on disordered polycrystalline superconductors~\cite{esquinazi1991, esquinazi1986}. From these $\alpha_L$ values, we estimate an average single-vortex pinning stiffness $\langle k_p \rangle = \alpha_{\mathrm{L}} \Phi_0/\Bip \approx 10$--$20\,\mathrm{N\,m^{-2}}$, consistent with defect-mediated pinning at grain boundaries and edges in polycrystalline thin-film nanostructures.

Superimposed on the smooth elastic background, we observe sharp, discrete frequency jumps, which we attribute to the stochastic entry of individual flux quanta or small vortex bundles, consistent with flux penetration dynamics observed in Ref.~\cite{bolle1999observation}. These features enable us to transcend ensemble averaging and directly quantify the mechanical impact of single-flux-quantum events. For small frequency shifts $\Delta \Omega = \Omega-\Omega_0$ at jumps, we can use the approximation $\Omega^2 - \Omega_0^2 \approx 2\Omega_0\Delta\Omega$ and obtain
\begin{equation}
\Delta\alpha_\mathrm{L} = 2\,\rho_\mathrm{eff}\,\Omega_0\,\Delta\Omega,
\label{eq:delta_labusch}
\end{equation}
enabling the direct extraction of Labusch stiffness changes from the data. Using the measured shifts of $\Delta\Omega/2\pi = 10$–$50$ Hz at jumps and $\Omega_0/2\pi \approx 5.8$\,MHz, we obtain collective stiffness changes of $\Delta\alpha_\mathrm{L} \simeq (4.2\text{–}21)\times 10^{12}\ \mathrm{N\,m^{-4}}$. Interpreting one resolved frequency jump as originating from the entry of a single vortex into the nanostring area ($A=ld\simeq 3.3\times10^{-12}$ m$^2$) gives the pinning stiffness per unit length $k_p \approx A\Delta\alpha_\mathrm{L} \approx 14$--$70~\mathrm{N\,m^{-2}}$. Assuming an optimum pinning site with radius $r_p\sim\xi\sim\SI{0.1}{\micro\meter}$ and vortex line length $L_\mathrm{line}\sim w\approx\SI{0.2}{\micro\meter}$, the corresponding pinning energy is $U\simeq\tfrac{1}{2}k_p r_p^2 L_\mathrm{line}\approx 0.09$--$0.44$~eV$\gg k_\mathrm{B}T$. For completeness, the associated increment of the mechanical spring constant is $\Delta k = 2m_\mathrm{eff}\Omega_0\Delta\Omega\approx (2.7\!-\!13)\times10^{-6} \mathrm{N\,m^{-1}}$. At the thermal rms displacement $x_\mathrm{rms}=\sqrt{k_{\mathrm{B}} T/(m_\mathrm{eff}\Omega_\mathrm{m}^2)}\approx\SI{1}{\pico\meter}$, the corresponding force variation is $\Delta F=\Delta k\,x_\mathrm{rms}\approx3-17\ \mathrm{aN}$. The thermal force noise sets the baseline sensitivity. In the classical limit $(k_{\mathrm{B}} T \gg \hbar\Omega_\mathrm{m})$, the double-sided force-noise power spectral density~\cite{gorodetksy2010determination} $S_{\mathrm{FF}}^{\mathrm{th}}=2\,m_{\mathrm{eff}}\Gamma_{\mathrm{m}}k_{\mathrm{B}}T$ gives $S_{\mathrm{FF}}^{\mathrm{th}}\approx 1.77\times10^{-37}$ N$^2$ Hz$^{-1}$ ($\sqrt{S_{\mathrm{FF}}^{\mathrm{th}}}\approx 0.42$ aN Hz$^{-1/2}$) for our device. Over a $1$--$10$ Hz bandwidth, $F_{\mathrm{rms}}\approx 0.42$--$1.3$ aN, well below the 3--17~aN force steps, confirming that individual vortex-entry events are mechanically resolvable. Unlike vibrating-reed techniques that average over many vortices, or scanning probes~\cite{auslaender2009,embon2015probing} that require applied piconewton (pN) forces, our passive, device-integrated approach captures spontaneous flux dynamics in real time, providing direct access to discrete vortex-entry processes and enabling an in situ mechanical determination of a device-specific first-entry field $B_\mathrm{c1}$ and upper critical field $B_\mathrm{c2}$ in extended field measurements.

\section{Conclusion}

We have demonstrated that cavity optomechanics provides a highly sensitive probe of vortex matter in mesoscopic superconductors. By embedding a superconducting aluminum nanostring into a SQUID-terminated microwave cavity and monitoring its mechanical resonance frequency, we uncover two distinct magnetic-field-dependent signatures that reveal complementary aspects of vortex physics. First, the resonance frequency exhibits a smooth power-law scaling, $\Omega_0^2 \propto \Bip^{1.81}$, consistent with elastic stiffening in the Campbell regime as vortex density increases. From this collective behavior, we extract a Labusch parameter of $\alpha_{\mathrm{L}} \sim 10^{14}~\mathrm{N\,m^{-4}}$, quantifying the ensemble-averaged pinning elasticity characteristic of disorder-mediated collective pinning in polycrystalline thin films. Superimposed on this background, discrete frequency jumps mark signatures suggesting individual Abrikosov vortex entry events, resolved with attonewton-scale force sensitivity and corresponding to single-vortex pinning energies of 0.1–0.4~eV. These vortex entry events occur at stochastically varying field values between thermal cycles, directly mapping the disorder landscape that governs nucleation pathways.

These results pave the way for using cavity optomechanics as a quantitative tool to investigate vortex physics and test pinning theories in mesoscopic superconducting structures. By operating directly in chip-relevant geometries, this approach enables the study of fundamental superconducting phenomena in architectures akin to those used for superconducting quantum circuits, and offers a route to identifying how material disorder and device design impact flux-noise–related decoherence. In addition, these findings advance cavity optomechanics itself—particularly inductive coupling schemes—by providing key guidance for the design of architectures that aim to operate in the single-photon strong-coupling regime.

\begin{acknowledgments}
We thank Akashdeep Kamra for stimulating discussions. We acknowledge funding by the Horizon Europe 2021-2027 Framework Programme under the grant agreement No 101080143 (SuperMeQ), from the Deutsche Forschungsgemeinschaft (DFG, German Research Foundation) under Germany’s Excellence Strategy—EXC-2111-390814868. This research is part of the Munich Quantum Valley, which the Bavarian state government supports with funds
from the Hightech Agenda Bayern Plus.
\end{acknowledgments}

\section*{Data Availability}
The data that support the findings of this study are openly available in Zenodo at the following DOI: 10.5281/zenodo.17629487.

\section*{Author Contributions}
T.L. and L.N. performed the experiments. H.H., T.L., and A.M. built the measurement setup. T.S.P. performed data analysis and wrote the manuscript with H.H., T.L., and R.G. R.G. and H.H. coordinated the project and contributed to discussions.

\section*{Conflict of Interest}
The authors declare no competing interests.

\appendix

\section{Modeling and Fitting Methodology}
\label{app:Fitting}
The flux-dependent mechanical frequency is obtained by fitting the data to Eq.~\eqref{fre} for each fixed in-plane field $\Bip$. Possible junction asymmetry is described by $\alpha$ (default $\alpha=0$). Fits are initialized with a critical current $I_0=\SI{1.3}{\micro\ampere}$, a loop inductance $L_\mathrm{l}=\SI{136}{\pico\henry}$, a mode-shape factor $\eta=0.9$, and a mechanical frequency  $\Omega_0/2\pi=\SI{5.8}{\mega\hertz}$. Moreover, we apply a flux offset \(\phi_\mathrm{off}\) in the Josephson-inductance term (\(\phi\!\to\!\phi+\phi_\mathrm{off}\), with \(\phi=\Phib/\Phi_0\)) to account for residual static fields. The fitted parameters are obtained by nonlinear least-squares optimization. For our previous publication~\cite{luschmann2022mechanical}, the fitting analysis was based on the model of Ref.~\cite{shevchuk2017},
\begin{equation}
\Omega_\mathrm{m}
= \sqrt{\Omega_0^2(\Bip) +
\frac{4 E_J \pi^2 \Bip^2 l^2 \eta^2 (1-\alpha^2)\,\mathcal{T}(\varphi_b,\alpha)}
{m_\mathrm{eff}\,S_0^3(\Phib)}}\,,
\label{eq:prev_compact}
\end{equation}
with $\mathcal{T}(\varphi_b,\alpha)= \cos^{4}\!\varphi_b-\alpha^{2}\sin^{4}\!\varphi_b$ and $\varphi_b= \pi\Phib/\Phi_0$. This model neglects the geometric loop inductance $L_\mathrm{l}$, which is explicitly taken into account in the present work through Eq.~\eqref{fre} for higher precision.

%

\end{document}